\documentclass{SciPost}

\hypersetup{
    colorlinks,
    linkcolor={red!50!black},
    citecolor={blue!50!black},
    urlcolor={blue!80!black}
}
\usepackage{subfigure}
\usepackage[subfigure]{tocloft}
\usepackage[bitstream-charter]{mathdesign}
\DeclareSymbolFont{usualmathcal}{OMS}{cmsy}{m}{n}
\DeclareSymbolFontAlphabet{\mathcal}{usualmathcal}

\fancypagestyle{SPstyle}{
\fancyhf{}
\lhead{\colorbox{scipostblue}{\bf \color{white} ~SciPost Physics }}
\rhead{{\bf \color{scipostdeepblue} ~Submission }}

\fancyfoot[C]{\textbf{\thepage}}
}

\begin{document}

\pagestyle{SPstyle}

\begin{center}{\Large \textbf{\color{scipostdeepblue}{
Reducing simulation-related emissions in rare-event searches through optimised event biasing \\
}}}\end{center}

\begin{center}\textbf{
Patrick Knights\textsuperscript{1},
Lachlan J. Milligan\textsuperscript{1$\star$} and
Konstantinos Nikolopoulos\textsuperscript{1,2}
}\end{center}

\begin{center}
{\bf 1} University of Birmingham, School of Physics and Astronomy, Birmingham, UK
\\
{\bf 2} University of Hamburg, Institute for Experimental Physics, Hamburg, Germany
\\[\baselineskip]
$\star$ \href{mailto:email1}{\small l.j.milligan@bham.ac.uk}
\end{center}

\section*{\color{scipostdeepblue}{Abstract}}
\textbf{\boldmath{%
Rare-event search experiments continue to extend their sensitivities to unprecedented levels. 
This requires increasingly small backgrounds, and shielding schemes that can suppress external backgrounds by several orders of magnitude. 
Moreover, detailed simulations are required to attain a good understanding of these experimental backgrounds. 
Studies have suggested, however, that simulations and computing-related tasks contribute approximately 10\% to the average particle physicist's carbon footprint. 
With backgrounds frequently below 0.01 counts per kg of target per keV of energy, simulations require more computing resources; growing a rare-event researcher's computing-related carbon footprint. 
Event biasing is often applied in shielding simulations as a remedy, yet there is a lack of systematic guidance on how to maximally benefit from them. 
An optimisation study for the importance-splitting biasing technique is discussed, focused on balancing statistical precision with simulation CPU-time, and how this can benefit the average researcher's carbon footprint.
}}

\vspace{\baselineskip}

\noindent\textcolor{white!90!black}{%
\fbox{\parbox{0.975\linewidth}{%
\textcolor{white!40!black}{\begin{tabular}{lr}%
  \begin{minipage}{0.6\textwidth}%
    {\small Copyright attribution to authors. \newline
    This work is a submission to SciPost Physics. \newline
    License information to appear upon publication. \newline
    Publication information to appear upon publication.}
  \end{minipage} & \begin{minipage}{0.4\textwidth}
    {\small Received Date \newline Accepted Date \newline Published Date}%
  \end{minipage}
\end{tabular}}
}}
}

\vspace{10pt}
\noindent\rule{\textwidth}{1pt}
\tableofcontents
\noindent\rule{\textwidth}{1pt}
\vspace{10pt}

\section{Introduction}
\label{sec:intro}
Rare event searches, growing in prominence within the field, are increasingly probing regions of parameter space that demand extremely low and well-understood experimental backgrounds. 
The direct detection of WIMP-like dark matter is one example, with searches at two frontiers: i) a high-mass but low cross-section frontier requiring ever-lower backgrounds for higher sensitivities; and ii) a low-mass frontier that requires a low energy threshold and thus low backgrounds in the same domain~\cite{Billard:2021uyg}. 
Consequently, high-fidelity and low-uncertainty background simulations are essential. 
A problem arises from the confluence of these two requirements, however, which is that background and shielding simulations with high statistics and low uncertainty are hard to achieve when the experiments themselves are designed to be ultra-low background. 

Event biasing schemes are a strategy to address this challenge, acting to effectively increase the statistical precision of a simulation by applying appropriate weights to biasing-generated events so to preserve the final `results' or outcome of the simulation. 
We focus on one such scheme implemented within \textsc{Geant4}~\cite{GEANT4Developments}, named importance splitting and Russian roulette~\cite{Kahn1956}. 
Briefly this method is summarised as the continual ``splitting'' of particles selected for biasing at the boundaries of layers defined within a detector geometry.

Whilst resources and studies exist for the implementation and supplementation of this technique in \textsc{Geant4}~\cite{SuperCDMSSplitting}, there is little in the way of detailed guidance regarding how to best simultaneously optimise the uncertainty and computational efficiency of a simulation. 
In this proceedings, we present results from an optimisation study of this biasing technique, alongside said study's implications regarding computational efficiency and, consequently, a reduced computing-related carbon footprint. 

\section{Optimal Importance Splitting/Russian Roulette Biasing}

\begin{figure*}[!h]
   \centering
   \subfigure[\label{fig:GammaRETrend_a}]{\includegraphics[width=0.465\textwidth]{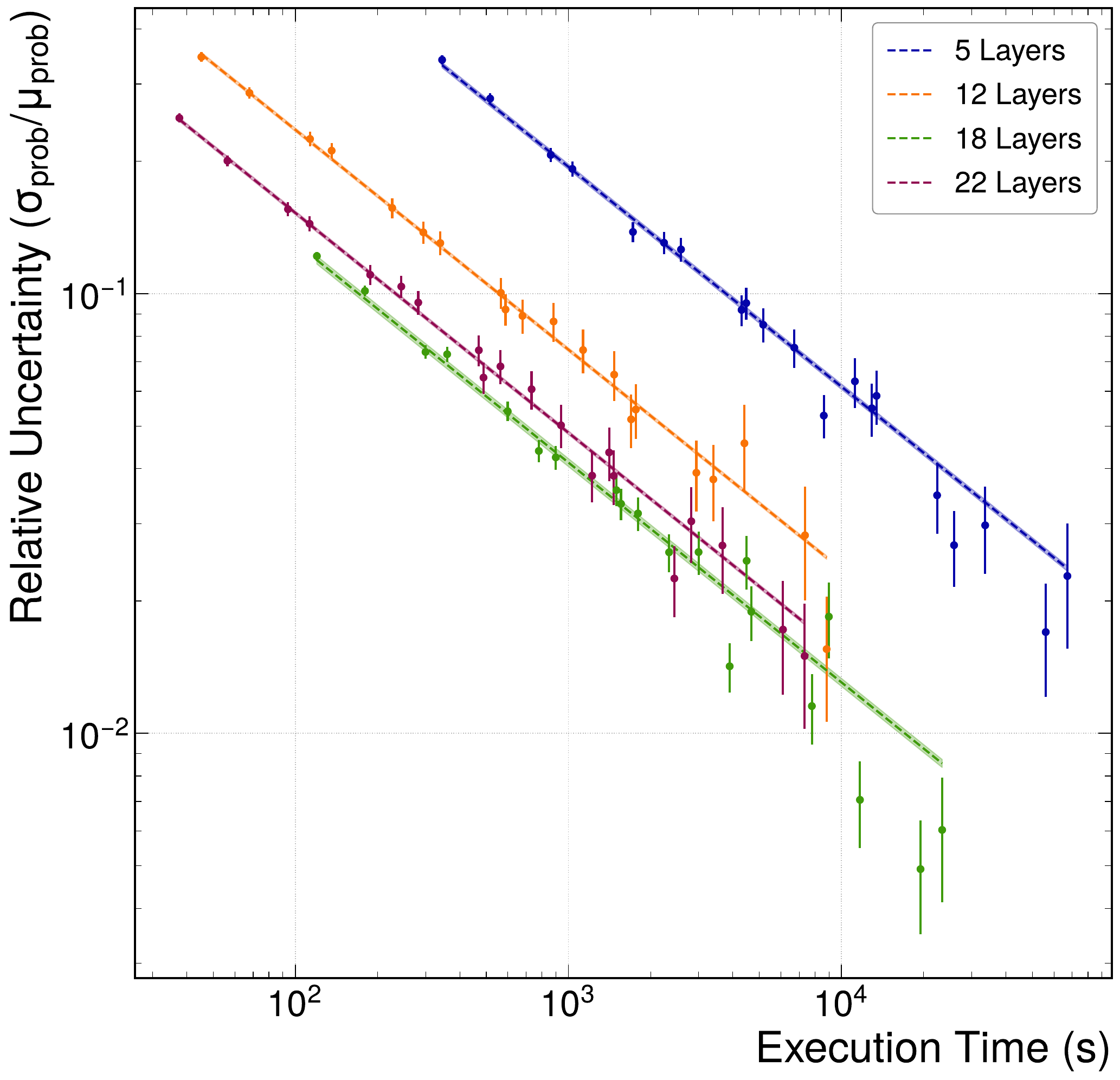}}
   \subfigure[\label{fig:GammaRETrend_b}]{\includegraphics[width=0.465\textwidth]{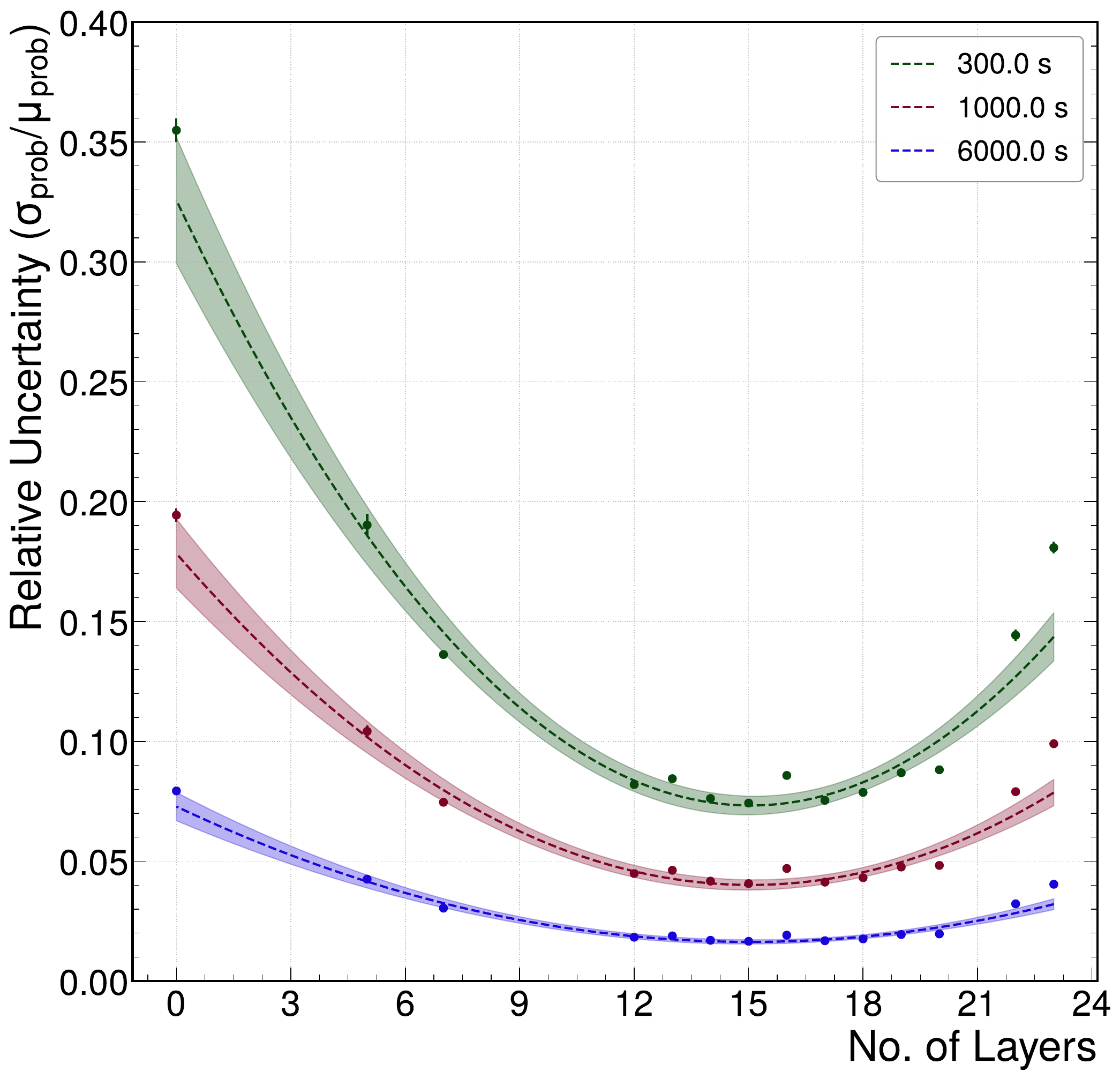}}
   \caption{\subref{fig:GammaRETrend_a} $\mathcal{R}$ versus execution time for different number of layers. \subref{fig:GammaRETrend_b} The observed performance trend. Shaded regions represent the $\pm 1\sigma$ fit uncertainty.
   \label{fig:GammaRETrend}}
\end{figure*}

For our study, we utilise a detector and shielding geometry inspired by the proposed DarkSPHERE experiment~\cite{DarkSPHERE:2023qwh} searching for sub-GeV particle-dark matter with a spherical proportional counter~\cite{NEWS-G:2024jms,Nikolopoulos:2025kws}. Specifically, we study the optimal number of equally thick importance layers within the 1.5~m thick shielding geometry, balancing execution time with relative uncertainty. 
To do so, we run simulations and combine the results into a series of datasets with increasingly larger amounts of primary events, and plot the results' relative uncertainty ($\mathcal{R}$) against simulation execution time. The resulting curves are fit 
with the function $\mathcal{R} = A \times t^{1/2}$, where $t$ is the execution time and $A$ the overall scale. This process is repeated for a number of different layer counts, for both $\gamma$-rays and neutrons at two energies each. For the purpose of this proceedings, we focus on $\gamma$-rays at 2750~keV. Figure~\ref{fig:GammaRETrend} shows an example of some fits for 2750~keV $\gamma$-rays. 

To see the full trend, we take a constant execution time and extract the respective value for $\mathcal{R}$ from each fit corresponding, and plot said value against layer count. Figure~\ref{fig:GammaRETrend} shows the result of this, which is fit to a parabola. The downward trend on the right is explained by the additional splitting increasing the available statistics, whilst the upward trend, i.e. worsening performance, can be explained by the computational burden of the split particles now outweighing the improvement in statistics. Given, therefore, the minimum of the parabola can be considered the optimal number of layers, we can utilise it as our point of comparison for improved CO$_2$e emissions. This methodology is explored in more depth, and also for 1461~keV $\gamma$-rays and neutrons at 10~MeV and 5~MeV, in Ref.~\cite{KnightsMilliganNikolopoulos}. In the following section we estimate the carbon footprint for different biasing scenarios, and perform the aforementioned comparison.

\section{Reducing Computing-related Greenhouse Gas Emissions with Optimal Biasing}
A large fraction of a researcher's professional carbon footprint, particularly if their work is predominantly computational, is often due to computing related carbon emissions. On average, for a researcher working at an LHC experiment, it is the third highest contributor to their total footprint~\cite{CO2eHEP:2024wpo}. Consequently, using benchmark efficiency improvements from the optimised event biasing, we can determine the corresponding reduction in computing-related CO$_2$e footprint. To do so, we use the methodology presented in Ref.~\cite{CO2eHEP:2024wpo} applied to the local cluster at the University of Birmingham (UoB). The expression for total computing emissions footprint, due to a simulation submitted to a high performance computing (HPC) cluster, is 
\begin{equation}
    {\rm Footprint~[gCO_2e]} = f_{\rm PUE} \times n_{\rm WPC} \times f_{\rm conv},
\end{equation}
where $f_{\rm PUE}$ is defined as the power usage effectiveness of the cluster, i.e. a quantification of how much of the cluster's total power usage is towards actual computation, $n_{\rm WPC}$ is the clusters workload power consumption (WPC), and $f_{\rm conv}$ is the conversion factor from kWh to gCO$_2$e, which is a factor dependent on the particular power grid you are on, and its balance of power sources. The workload power consumption is defined separately as:
\begin{equation}
    n_{\rm WPC} = p_{\rm CPU-core} \times t_{\rm core-h},
\end{equation}
where $p_{\rm CPU-core}$ is the power consumption of a one CPU core in kW, and $t_{\rm core-h}$ is the time it was under load. All these parameters, except $t_{\rm core-h}$, are constant as they are cluster specific. 

For the local cluster at UoB $f_{\rm PUE}=1.5$, which is the global average~\cite{CO2eHEP:2024wpo}, and we set $f_{\rm conv}=174$~gCO$_2$e/kWh, which is the UK's average for 2025~\cite{elecmaps}. The cluster is equipped with AMD EPYC 7513 32-Core Processors with hyper-threading, which has a maximum overall power draw of 200~W or 6.25~W per core. Assuming the maximum power draw, to be conservative, we set $p_{\rm CPU-core}=6.25$~W. The execution time is used as a proxy for $t_{\rm core-h}$ in this case, given the execution time represents the CPU-time of the core program.

Combining the constant parameters, the expression for the UoB cluster becomes:
\begin{equation}
    {\rm Footprint~[gCO_2e]} = 1.63 \times t_{\rm core-h}.
\end{equation}

We use the optimal point in Fig.~\ref{fig:GammaRETrend} as our basis for comparison, which corresponds to 15 layers with $\mathcal{R}=0.074$ in 300.0~s. We compare three non-optimal cases to this point: an under-biased (5 layers), over-biased (23 layers), and unbiased simulation. We extract the respective times for each from their fits, and compute the resulting emissions footprint alongside the factor increase on the optimal. These results are shown in Table~\ref{tab:emissions}.
\begin{table}[htb]
\centering
\caption{Summary of emissions and their increase with respect to the optimal biasing case. Emissions are estimated for a single CPU core.}
\begin{tabular}{l|ll}
Biasing Level & Footprint  [gCO$_2$e] & Factor Increase on Optimal\\
\hline
Optimal (15L) & $0.135$ & -\\
Under-biased (5L) & $0.890$ & $6.6$ \\
Over-biased (23L) & $0.804$ & $5.0$ \\
Unbiased & $3.099$ & $23.0  $ \\
\hline
\end{tabular}
\label{tab:emissions}
\end{table}

Applying this to a proposed experiment, we can get an idea of how the reduction in carbon footprint looks in practice. We use the DarkSPHERE detector and shielding geometry~\cite{DarkSPHERE:2023qwh}, producing primary $\gamma$-rays at 2750~keV isotropically from the surface of the 2.5~m thick shielding. Similar to the results in Table~\ref{tab:emissions}, we compare the execution times required to reach the value for $\mathcal{R}$ derived from the most optimal layer configuration. The optimal layer configuration is 25 layers, and we compare against the unbiased case. We find that, using 25 layers, a value of $\mathcal{R}=0.05$ can be achieved in $t_{\rm core-h}=0.83$~h, whilst for the unbiased case the same $\mathcal{R}$ is achieved in $t_{\rm core-h}=841.5$~h . This translates into $1.4$~gCO$_2$e and $1371.6$~gCO$_2$e, respectively, i.e. a reduction of almost three-orders of magnitude. The unbiased footprint also approximates emissions for visiting the experiment site (Boulby Underground Laboratory) from Birmingham (approx. 14~kgCO$_2$e).

Two additional variables can also lead to a reduced footprint: $f_{\rm conv}$ and $f_{\rm PUE}$. They represent the CO$_2$e produced per kWh of power by the researcher's home grid and the power usage effectiveness of the HPC cluster used, respectively. For example, France has a value of $f_{\rm conv} = 32$~gCO$_2$e/kWh, which is significantly lower than the UK. Similarly, clusters exist with improved values for $f_{\rm PUE}$, due to better cooling etc. Thus, the best way for an individual researcher to reduce their simulation-related footprint, would be to utilise an optimised biasing scheme, and to perform the computations on a modern cluster with less power/cooling overheads located within a grid with a high proportion of clean energy. The greatest reduction is always achieved by combining all three strategies.

\section{Conclusion}
It is becoming more common for low background rare-event searches to utilise event biasing in their background simulation studies. A commonly used biasing technique is importance layer splitting and Russian roulette. Whilst effective, there is a lack of empirical advice for its optimisation, particularly with regards to the optimal layer thickness. We show an example of such an optimisation.
In using the optimal layer configuration, one can significantly reduce CPU resources required and, consequently, their simulation-related carbon footprint. We see a reduction in the carbon footprint for the the optimally biased simulation when compared to the under- and over-biased cases, and a substantial reduction compared to the unbiased scenario. In a practical application we see this reduction drastically increase, indicating benefits increase for thicker shielding designs. It is evident, therefore, that not only statistical and computational gains can be made from implementing optimal biasing schemes, but also substantial reductions to a rare-event researchers computational-related carbon footprint.

\section*{Acknowledgements}
The authors acknowledge the support of the UKRI-STFC (ST/X005976/1, UKRI/ST/C002848/1, ST/W000652/1), and of the Deutsche Forschungsgemeinschaft (DFG, German Research Foundation) under Germany’s Excellence Strategy – EXC 2121 “Quantum Universe”-390833306.

\bibliography{bibtex.bib}

\end{document}